\documentclass[a4paper,12pt]{article}

\usepackage{amsmath}        
\usepackage{amsfonts}       
\usepackage{amssymb}        
\usepackage{graphicx}       
\usepackage{subcaption}     
\usepackage{caption}
\usepackage{xcolor}
\usepackage{booktabs}
\usepackage{hyperref}
\usepackage{footmisc}
\usepackage{appendix}
\usepackage[margin=2cm]{geometry}

\begin{document}

\title{Statistical permutation quantifiers in the classical transition of conservative-dissipative systems}

\author{ Gaspar Gonzalez$^{1,2}$\footnote{Corresponding author: \href{mailto:autor@example.com}{gaspar.gonzalez@fisica.unlp.edu.ar}} \and Andrés M. Kowalski$^{1,2}$  \\
    \textit{$^{1}$Instituto de Física de La Plata CONICET/UNLP,  Buenos Aires, Argentina}\\
    \textit{$^{2}$ Comisión de Investigaciones Científicas (CIC), Argentina}
}
\date{\today}

\maketitle

\begin{abstract}
    We study the behavior of a nonlinear semiclassical system using Shannon entropy and two approaches to statistical complexity. These systems involve the interaction between classical variables (representing the environment) and quantum ones. Both conservative and dissipative regimes are explored. To calculate the information metrics, probability distributions are derived from the temporal evolution via the Bandt-Pompe permutation method. Additionally, we describe the classical limit in terms of a motion invariant linked to the uncertainty principle. Our analysis reveals three distinct regions, including a mesoscopic one, along with other notable findings.
\end{abstract}

\section{Introduction}
The emergence of the classical world from the quantum domain is a topic of significant interest, both theoretically \cite{Struyve,Brack,Arndt} and experimentally \cite{Li,Goan,Barker}, as well as for its practical applications. The decoherence process \cite{Joos}, for instance, is closely linked to mesoscopic physics a field focusing on materials of intermediate scales. Mesoscopic systems, such as those used in nanotechnology, serve as practical examples \cite{das2010mesoscopic, brandes2005coherent, x12, x13}. Additionally, semiclassical systems have been extensively employed over the years to address various physical problems \cite{Bloch,Milonni,nielsen2001stat,micklitz2013semicla, cosme2014the,prants2017quantum,nam2,x2,x3,x4,x5,x6}.

We have investigated systems that exhibit the coexistence of quantum and classical variables \cite{ref, AK1, Kowalski2002, RK1, AK2, KPG21, Chaos}. In these systems, the classical variables serve as a representation of the reservoir interacting with the quantum system.

Within this framework, we have explored the classical limit of quantum systems by employing a motion invariant associated with the Heisenberg Principle. Our approach includes dynamic analyses \cite{ref, AK1, Kowalski2002, RK1, AK2, Chaos} and the application of various tools from Information Theory (IT) \cite{KMBP07,KMP15}.

The classical limit of the system addressed in this work was previously examined in Ref. \cite{Chaos}, utilizing Poincaré sections and analyzing the temporal evolution of key quantities. Both conservative and, for the first time, dissipative regimes were considered. In this study, we revisit the problem by employing Shannon entropy and two definitions of statistical complexity: the LMC complexity \cite{LR1995} and the Jensen-Shannon complexity \cite{MPR}. To compute any information quantifier, a probability distribution function is required. For this purpose, we apply the Bandt-Pompe permutation based symbolic method \cite{BP2002}, recognized for its efficiency in capturing key features of various time series \cite{KS05, RZPF07, SCFR10, ROZM13, ACAF15, HJ19}.

Our objective is to gain a deeper understanding and characterization of the process, uncovering details that cannot be obtained using dynamic tools alone.

\section{The semi-classical model and  classical limit}
We consider a Hamiltonian of the form 
\begin{equation}
 \hat{H}= \frac{1}{2} \left(\omega_{q}\left(\hat{x}^2+\hat{p}^2\right)+ \omega_{cl}\left(A^{2}+P_{A}^{2}\right) \hat{I} +e_{q}^{cl} A^{2}\hat{x}^{2}\right).
\label{eq4}
\end{equation}
In this Hamiltonian, $\hat{x}$ and $\hat{p}$ are considered quantum operators while  $A$ and  $P_{A}$ canonically conjugated classic variables. $\hat{I}$ is the Identity operator. $\omega_{q}$ and $\omega_{cl}$ are frequencies and $e_{q}^{cl}$ a positive parameter positive \cite{ref}. 
The non-linear quantum-classical interaction appears through the term  $e_{q}^{cl} A^2 \hat{x}^2$ in (\ref{eq4}). 
The semiclassical methodology that we use, is developed in the Appendix. The equations of motion that govern the system, as demonstrated in Ref. \cite{Chaos}, are given by the following set of coupled non-linear equations

\begin{equation}
\begin{aligned}
\label{eq semi}
&\dfrac{d\langle \hat{x}^2\rangle}{dt} =\omega_{q} \langle \hat{L}\rangle, \quad  \dfrac{d\langle \hat{p}^2\rangle}{dt}  =-\left(\omega_{q} +e^{cl}_{q} A^{2}\right)\langle \hat{L}\rangle, \\
& \dfrac{d\left\langle \hat{L}\right\rangle}{dt}  = 2\left(\omega_{q} \left\langle\hat{p}^{2}\right\rangle-\left(\omega_{q}+e^{cl}_{q} A^{2}\right)\left\langle\hat{x}^{2}\right\rangle\right),\quad \hat{L}= \hat{x}\hat{p}+\hat{p}\hat{x} \\
&\dfrac{dA}{dt}=\omega_{cl} P_{A},\quad  \dfrac{dP_{A}}{dt}=-A\left(\omega_{cl}+e^{cl}_{q} \left\langle \hat{x}^{2}\right\rangle\right)-\eta P_{A}. 
\end{aligned}
\end{equation}

To study the classical limit of the system (\ref{eq semi}), we need to compare with the solutions corresponding to  the classical analogous of the Hamiltonian (\ref{eq4}) \cite{Kowalski2002} 
\begin{equation}
H=\frac{1}{2} \left(\omega_{q}\left(x^2+p^2\right)+\omega_{cl}\left(A^2+P_{A}^{2}\right)+e^{cl}_{q} A^{2}x^{2}\right).
\label{Hclasico}
\end{equation} 

By Hamilton's equations we obtain (see Appendix)
\begin{equation}
\label{eqTC}
\begin{aligned}
&\dfrac{dx^{2}}{dt} =\omega_{q} L,  \quad \dfrac{dp^{2}}{dt}=-\left(\omega_{q}+e_{q}^{cl} A^{2}\right)L,  \\
&\dfrac{dL}{dt}=2\left(\omega_{q} p^{2}-\left(\omega_{q}+e_{q}^{cl},  A^{2}\right)x^{2}\right), \quad L=2xp,\\
&\dfrac{dA}{dt}=\omega_{cl} P_{A},\quad
\dfrac{dP_{A}}{dt}=-A\left(\omega_{cl}+e_{q}^{cl}x^{2}\right)-\eta P_{A}.
\end{aligned}
\end{equation}

We define   $I$  as 
\begin{equation}
I=\langle \hat{x}^{2}\rangle \langle \hat{p}^{2}\rangle -\dfrac{\langle \hat L\rangle ^{2}}{4} \geq \dfrac{\hbar ^{2}}{4}.
\label{eqInv}
\end{equation}
 $I$  is a motion-invariant for the systems (\ref{eq semi})  in both regimens (conservative and dissipative) (see \cite{ref,Chaos}) and is related to the uncertainty principle describing the deviation with respect to classicity, a condition characterized by $I={x}^2 \, {p}^2-  L^2 /4=0$ (trivial motion-invariant).\\
Eqs. (\ref{eq semi}) do not  explicitly depend upon  $\hbar$ while mean values depend on it via the initial conditions  through $I$ satisfying (\ref{eqInv}). The inequality (\ref{eqInv}) and its classical value, are crucial in our study of the classical limit $I \rightarrow 0$, as will be seen throughout the article.
Focus now attention upon the relative energy  $E_{r}$. 
\begin{equation}
E_{r}=\dfrac{|E|}{I^{1/2}\omega_{q}},
\label{eq16}
\end{equation}
where $E=\left\langle H \right\rangle$ is the total energy. $E_{r}$ is also a motion invariant that verifies $E_{r}\geq 1 $ due to the uncertainty principle. $E_{r}= 1$ corresponds to the minimum value and represents the totally quantum case. $E_{r}$ is dimensionless. The LC is  $I \rightarrow 0$, i.e., 
\begin{equation}
E_{r} \to \infty.
\label{eq16infty}
\end{equation}
 In the dissipative case we replace, in  (\ref{eq16}), $E$ by $E(0)$  \cite{Chaos}. 
Our main interest lies in the evolution of observable when $E_r$ grows from its minimum unity value $E_r=1$.

\section{Bandt-Pompe method}
To calculate our Information quantifiers, it is necessary to have a probability distribution associated with the data series that one wants to analyze. In this case, of a time series of the semiclassical and classical systems.
We obtain this probability distribution function (PDF) using the Bandt and Pompe method \cite{BP2002}.  This methodology basically consists of a symbolization process that is described as follows: Let $\{x_{t}\}_{t=1, 2, \cdots, N}$ be an arbitrary time series, the first step is to divide the time series into $n=N-\left(d-1\right)\tau$ overlapping partitions comprising $d > 1$ observations separated by $\tau\geq 1$ time units. For given values of $d$ and $\tau$, each data partition can be represented by
\begin{equation}
w_{p} = \left(x_{p}, x_{p+\tau}, x_{p+2\tau}, \cdots, x_{p+(d - 2)\tau}, x_{p+(d- 1)\tau}\right),
\end{equation}
where $p=1, 2, \cdots, n$ is the partition index.
The parameters $d$ and $\tau$ are the so-called embedding dimension and the embedding delay, respectively, the only two parameters of the Bandt and Pompe method. The original proposal presented by Bandt and Pompe was limited to $\tau = 1$, this means that the data partitions comprised consecutive elements of time series, however, Cao et al. \cite{Cao}, Zunino et al. \cite{ZSFRM10} and Pessa and Ribeiro \cite{ordpy}, studied the embedding delay for values greater than 1.

Then, for each partition $w_p$ we evaluate the permutation $\pi_p=\left(r_0,r_1, \cdots,r_{d-1}\right)$ of the index numbers $\left(0, 1, \cdots, d - 1\right)$ which arranges the elements of $w_p$ in ascending order, that is, the permutation of the index numbers defined by the inequality $x_{p+r_0} \leq x_{p+r_1} \leq - -- \leq x_{p+r_{d-1}}$. In case of equality of values, the order of appearance of the elements of the partition is maintained, i.e., if $x_{p+r_{k-1}} = x_{p+rk}$, then $r_{k -1} < r_{k}$ for $k = 1, \cdots, d - 1$.
After evaluating the permutation symbols associated with all data partitions, a symbolic sequence $\{\pi_{p}\}_{p=1,2,\cdots, n}$ is obtained.

The ordinal probability distribution $P=\{p_{i}\left(\Pi_{i}\right)\}_{i=1,2,\cdots, n_\pi}$ is the relative frequency of all possible permutations within the symbolic sequence
\begin{equation}
    p_{i}\left(\Pi_{i}\right)=\dfrac{\textrm{number of partitions of type } \Pi_{i} \, \textrm{in } \{\pi_{p}\}}{n}.
\end{equation}
Here $\Pi_{i}$ represents each of the $n_{\pi}=d!$ different ordinal patterns.

The Bandt Pompe method has desirable features (i) simplicity, (ii) extremely fast calculation-process, and (iii) robustness. It is also invariant with respect to non-linear monotonous transformations. This method can be applied to any type of time series (regular, chaotic, noisy, or experimental) \cite{KMBP07}.

\section{Information Quantifiers}
\subsection{Shannon permutation entropy}
\label{Shannon_entropy}
The Shannon permutation entropy is defined as follows
\begin{equation}
    S\left(P\right)=-\sum\limits_{i=1}^{n_{\pi}} p_{i}\left(\Pi_{i}\right) \log{p_{i}\left(\Pi_{i}\right)},
    \label{shannon}
\end{equation}
where the probability distribution  $P=\{p_{i}\left(\Pi_{i}\right)\}_{i=1,2,\cdots, n_\pi}$  is obtained using the Bandt and Pompe method. 
Permutation entropy quantifies the randomness in the ordering dynamics of a time series. This means that $S= 0$ implies completely regular behavior, while $S=\log{\left( n_{\pi}\right)}$ indicates behavior with maximum disorder  \cite{BP2002, ordpy}. 

The normalized permutation entropy adopts  the form 
\begin{equation}
H\left(P\right)=\dfrac{S\left(P\right)}{S_{\textrm{max}}},
\label{shannonnormal}
\end{equation}
where $S_{\textrm{max}}=\log{\left(n_{\pi}\right)}$ y $0\leq H\leq 1$.

\subsection{LMC-statistical complexity}
\label{LMC_C}
In 1995, Lopez-Ruiz, Mancini and Calbet (LMC) \cite{LR1995, LR2001}, proposed a measure of complexity given by the following expression
\begin{equation}
    C=H\left(P\right)\cdot D\left(P, P_{e}\right),
\end{equation}
where $D\left(P, P_{e}\right)=\displaystyle\sum\limits_{i=1}^{n_{\pi}}\left(p_i-1/n_{\pi}\right) ^2$ is known as the disequilibrium factor and is defined as a distance (Euclidean or 2-norm distance) between a probability distribution $P$ and the equiprobable distribution $P_{e}$,  with  $P_{e}=\{1/n_{\pi}\}_{i=1,2,\cdots, n_\pi}$. Disequilibrium gives us a way to measure how  ``separate" the probability distribution of the system under treatment is from the uniform distribution, which gives us maximum uncertainty of the state of the system.

\subsection{Jensen Shannon-statistical complexity}
\label{JSC}
The Jensen Shannon-statistical complexity (JSC) \cite{MPR} is based on the pioneering Statistical complexity of López-Ruiz et al. The fundamental difference between these two quantities lies in how distances are measured in the probability space. The JSC is defined as the product of the permutation entropy and the Jensen-Shannon divergence between the ordinal distribution $P=\{p_{i}\left(\Pi_{i}\right)\}_{i=1 ,2,\cdots, n_\pi}$ and the uniform distribution $P_{e}=\{1/n_{\pi}\}_{i=1,2,\cdots, n_\pi}$ instead of Euclidean distance. We also consider the normalized version of the JSC.
\begin{equation}
C=\dfrac{H\left(P\right) \cdot D_{JS}\left(P, P_{e}\right) }{D^{\textrm{max}}},
\label{MPR-C}
\end{equation}
where 
\begin{equation}
D_{JS}\left(P, P_{e}\right)=S\left[ \dfrac{1}{2}\left(P+P_{e}\right)\right]-\frac{1}{2}  S\left(P\right) - \dfrac{1}{2} S\left(P_{e}\right),
\label{diverJS}
\end{equation} 
is the Jensen-Shannon divergence \cite{Lin} and 
 \begin{equation}
     D^{\textrm{max}}=-\dfrac{1}{2}\left[\dfrac{n_{\pi}!+1}{n_{\pi}!}\log\left(n_{\pi}!+1\right)-2\log\left( 2n_{\pi}!\right)+\log{\left(n_{\pi}!\right)}  \right],
     \label{ctenorm}
 \end{equation}
is a normalization constant.

The Jensen Shannon statistical complexity  is capable of capturing essential details of the dynamics, is an intensive quantity, and is capable of discerning between different degrees of periodicity and chaos \cite{ KMBP07, Lin, ZSR12,RLMP07}.

\section{Results}
To solve the equations (\ref{eq semi}) and (\ref{eqTC}) we choose the following parameter values: $\omega_q=\omega_{cl}=e_{q}^{cl}=1$. While for the initial conditions we take $\langle \hat{L}\rangle\left(0\right)=L(0)=A\left(0\right)=0$.

For the conservative system ($\eta=0$), we choose $E=0.6$. In the dissipative regimen we take the same value for the initial energy value $E(0)=0.6$. In both regimes we fixed this values, as we set $I\to 0$ ($E_{r}\to \infty$) in (\ref{eq16}). In the dissipative cases is $\eta=0.05$.

We select $\langle \hat{x}^{2}\rangle\left(0\right)=E/\omega_q- 0.98\sqrt{(E/\omega_q)^{2}-I_L}$, where $ I_L =I+\langle L\rangle (0)^{2}/4$. For the classic case is $x^{2}\left(0\right)=0.02E/\omega_q$, which results from taking the value $I=0$ in the previous expression. By appeal to (\ref{eqInv}) we see that 
 $\langle\hat{p}^{2}\rangle (0) =\dfrac{I}{\langle  \hat{x}^{2}\rangle} (0)$ ($p^2=0$ in the classical instance). In addition,  
 from  (\ref{eq4}) we deduce  $P_A (0)$ via
\begin{equation}
\label{PA0}
P_{A}= \frac{ \pm}{\omega_{cl}^{1/2}} \sqrt{2 E - \omega_q (\langle\hat{p}^{2}\rangle  +\langle\hat{x}^{2}\rangle) - e_{q}^{cl} A^2\langle\hat{x}^{2}\rangle - \omega_{cl} A^2 },  \nonumber \end{equation} 
evaluating all quantities at the initial time. We have also an equivalent expression for the classical analogous  system.

In the numerical calculation of the Shannon permutation entropy and the statistical complexities, the Python package ordpy developed by Arthur A. B. Pessa and Haroldo V. Ribeiro \cite{ordpy} was used. We have employed the values $d=5 $ and $\tau=1$ for the dimensional embedding and embedding delay, respectively.  The results have been conceptually confirmed using $d=6$.
 
The time series for the semiclassical case, have been constructed with the values of $\langle \hat{x}^{2}\rangle (t)$, each one for a different value of $E_r$ (or $I$). Also we have considered in the classical scenario ($I=0$), the unique series formed by the $x^2 (t)$  values. In both cases to we have taken $N=20000$ points for each series. The condition $ N>> d!=120$ is verified \cite{KMBP07}.

In all  figures we observe low values of the information quantifiers, unlike what we have found for the chaotic Hamiltonian of Refs. \cite{KMBP07,KMP15}. This feature are consistent with the regularity observed in the Poincaré Sections \cite{Chaos}. In detail, after the jump observed in the quasi-quantum zone between $E_r \approx 1$ and $E_r\approx 2.8$ (see below), one finds a very small band of variation of the values. However, the three regions of Refs. \cite{KMBP07, KMP15} of the path to the classical analogous, continues being observed (quasi-quantum, transitional and classical). 

It is convenient an analysis with additional tools such as information quantifiers, which, as mentioned, have proven efficiency for this type of problem.

In Fig.1 we plot the normalized Entropy $H$ vs. the relative energy $E_r$, for both conservative and dissipative regimes. In the conservative case (blue curve), between $E_r \approx 1$ and $E_r\approx 2.8$, we find out a zone, what we can call quasi-quantum. In the zone $E_r \approx 1$, with dynamic tools, quasi-periodic behavior has been observed (remember that $E_r = 1$ is the lowest possible value of $E_r$ and corresponds to the fully quantum case). At $E_r \approx 2.8$ (first gray colored vertical dashed line),  we have observed on a logarithmic scale, a drastic change (taking into account the mentioned small band of variation of $H$). The transition zone (mesoscopic) can be set between the values $E_r \approx 2.8$ and  $E_r \approx 104.8$ and finally the convergence zone to the classical analogous value $H_{c}^{cl}=0.1768$, can be seen from $E_r \approx 104.8$ (second gray discontinuous vertical line).  

The equivalent zones corresponding to the dissipative regime (orange curve) are: quasi-quantum from $E_r \approx 2.8$ to $E_r\approx 3.6$ (first plum colored vertical dashed line), the transitional from $E_r\approx 3.6$ to $E_r< 123.0$, and the convergence zone to the analogous classic case from $E_r=123.0$ (second plum colored vertical dashed line). 
$H_{d}^{cl}= 0.1720$, is the value of the entropy of the classical analogous in the dissipative case (remember that in this regime we replace in  (\ref{eq16}), $E$ by $E(0)$).

Additionally, the figure contains an inset in which the morphology of the curves in the transition zone is observed in detail. 
This region can be associated with a mesoscopic one, due to the important presence of quantum features. The zone of convergence to the totally classical system must correspond to a decoherence process \cite{KPG21}.

In Figs. \ref{CvsEr} we plot the two definitions of the statistical complexity vs. $E_r$. In both figures, one can clearly notice the three zones mentioned in Fig.1. These results are the same for both $C_{JS}$ and $C_{LMC}$. The blue curve corresponds to the conservative case and the orange curve to the dissipative one. The dashed vertical lines indicate the transition zone, gray for the conservative regime (between $2.8<E_r<104.8$) and plum for the dissipative case (from $E_r > 3.6$ to $E_r< 123.0$). Furthermore, in each graph there is a inset that basically consists of an increase in scale of the original graph, showing the transition zone. 
In Figure \ref{CvsEr_JS} we can see the statistical complexity of Jensen-Shannon $C_{JS}$. $C_{c}^{cl} =0.16401$ and $C_{d}^{cl}=0.16058 $ are the statistical complexity values of the classical analogous of the conservative and dissipative regimes, respectively. Figure \ref{CvsEr_cd_LMC} presents the statistical complexity introduced by Lopez-Ruiz et al. $C_{LMC}$ vs  $E_r$. The blue  and the orange curves are associated to the conservative and dissipative system, respectively. In this case, $C_{c}^{cl}=0.13533$  and $C_{d}^ {cl}=0.13426$ are the values of the statistical complexity of the classical analogous (conservative and dissipative regimen).

\begin{figure}[ht]
    \centering
   \includegraphics[height=7cm]{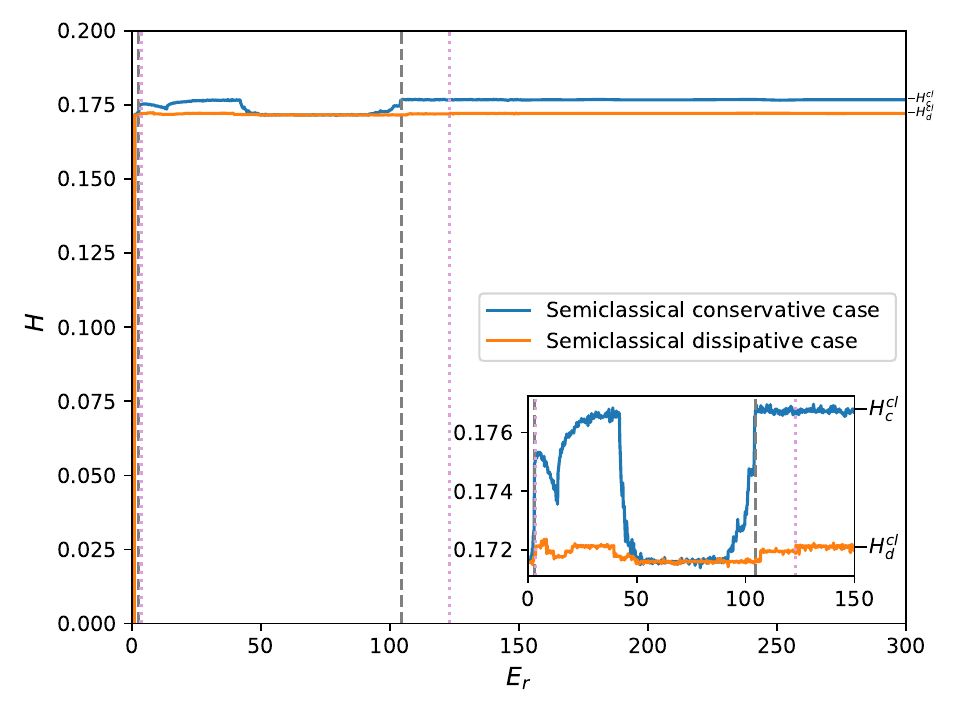}
   \caption{Normalized Shannon permutation entropy $H$ as a function of the relative energy of the system $E_{r}$. The insets on the right are an up-scaling of the original graph, showing the transitions zones with more details. The blue curve corresponds to the entropy of the conservative case an the orange to the dissipative one. $H_{c}^{cl}= 0.17676$ and $H_{d}^{cl}=0.17207$ are the corresponding entropies of the classical analogous in both regimes. Vertical lines indicate the transition zone, between quasi-quantum and classical zones. Dashed gray for the conservative system ($2.8<E_r<104.8$) and dotted plum for the dissipative one ($3.6 < E_{r_0 }< 123.0$).}
        \label{HvsEr}
\end{figure}

\begin{figure}
    \centering
      \begin{subfigure}{0.7\textwidth}
        \centering
        \includegraphics[height=7cm]{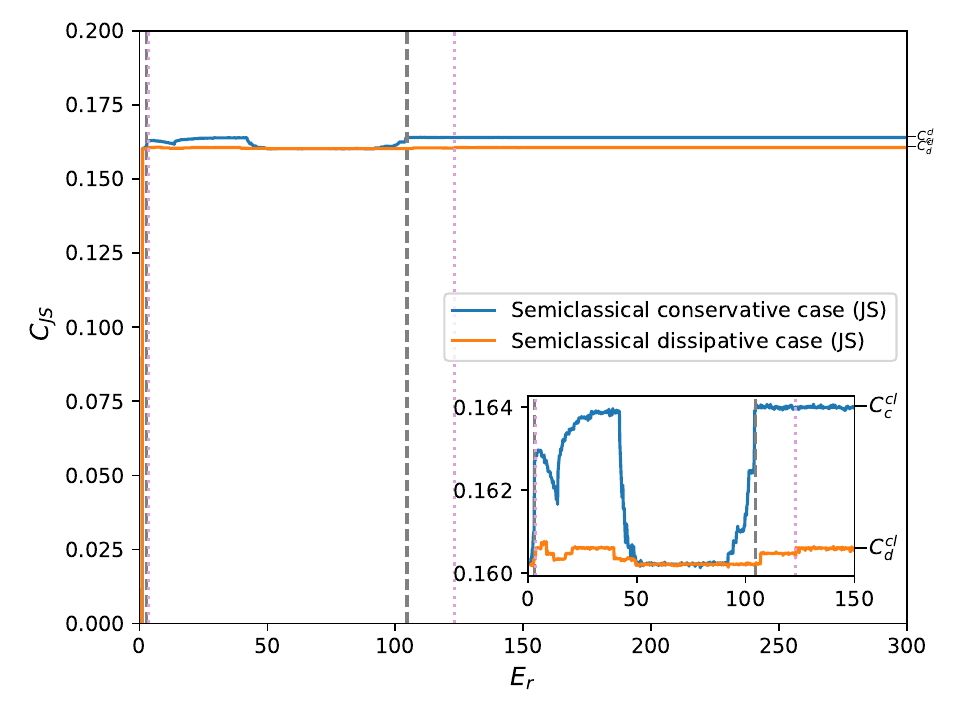}
       \caption{}
       \label{CvsEr_JS}
    \end{subfigure}
      \centering
      \begin{subfigure}{0.7\textwidth}
        \centering
        \includegraphics[height=7cm]{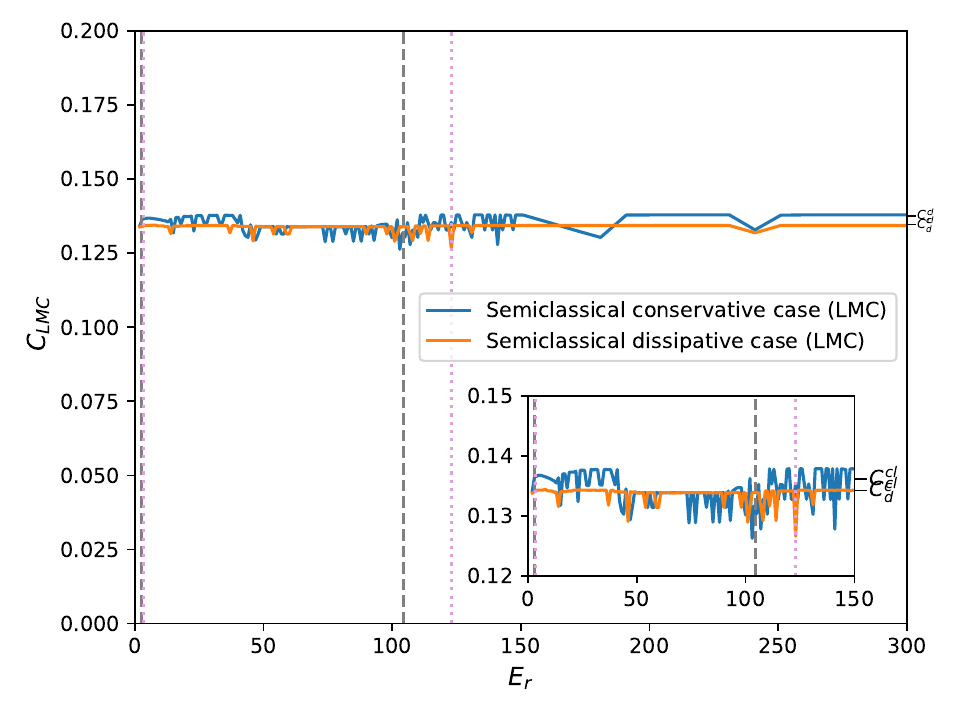}
       \caption{}
        \label{CvsEr_cd_LMC}
    \end{subfigure}
    \caption{ We plot the two definitions of the statistical complexity vs. $E_r$. As in previous figure, vertical lines indicate the transition zone, between quasi-quantum and classical zones. Dashed gray for the conservative dynamics ($2.8<E_r<104.8$) and dotted plum for the dissipative one ($3.6 < E_{r_0 }< 123.0$). The intervals are the same as in the case of entropy. In the insets we can see the transition zones in detail. The blue curves correspond to the conservative regime and the orange curves to the dissipative one.   
    (a) Statistical complexity $C_{JS}$ vs $E_r$.  $C_{c}^{cl}=0.16401$ and $C_{d}^{cl}=0.16058 $ are the statistical complexity values of the classical analogous in both conservative and dissipative cases, respectively. (b) Statistical complexity $C_{LMC}$ vs $E_r$. $C_{c}^{cl}=0.13533$ and $C_{d}^ {cl}=0.13426$ are the corresponding statistical complexity values of the classical analogous (conservative and  dissipative regimes).}
    \label{CvsEr}
\end{figure}

\clearpage
 \subsection{Comparing outcomes.}
 
As we have already commented, all figures show convergence towards the values of the corresponding classical analogous. Except for fluctuations, in general the results corresponding to the conservative regime are slightly higher than those of the dissipative regime. A result that can be expected since the dynamics of the system is regular in both regimes. However, contrary to what one might think a priori, convergence start earlier in the conservative case. The zones are the same for the entropy $H$ and the complexities $C_{JS}$ and $C_{LMC}$. Transitional zone of the conservative regime is ($2.8<E_r<104.8$), while for the dissipative case it is ($3.6<E_r<123.0$). 

Note that the representative figures of $H$ and $C_{JS}$ are similar but with a difference in scale. This is because the disequilibrium remains almost constant around its mean value $D_{JS} \approx 0.92985$ (conservative) and  $D_{JS} \approx 0.93337$ (dissipative),  for all values of $E_r$. In the case of the complexities $C_{LMC}$, the disequilibrium  is also quasi-constant as a function of $E_r$, but it has greater fluctuations around its mean value $D_{LMC\, c} \approx 0.76944$ and $D_{LMC\, d} \approx 0.77906$ (conservative and dissipative dynamics).

\begin{figure}[h]
    \centering
      \begin{subfigure}{0.49\textwidth}
        \centering
        \includegraphics[height=6cm]{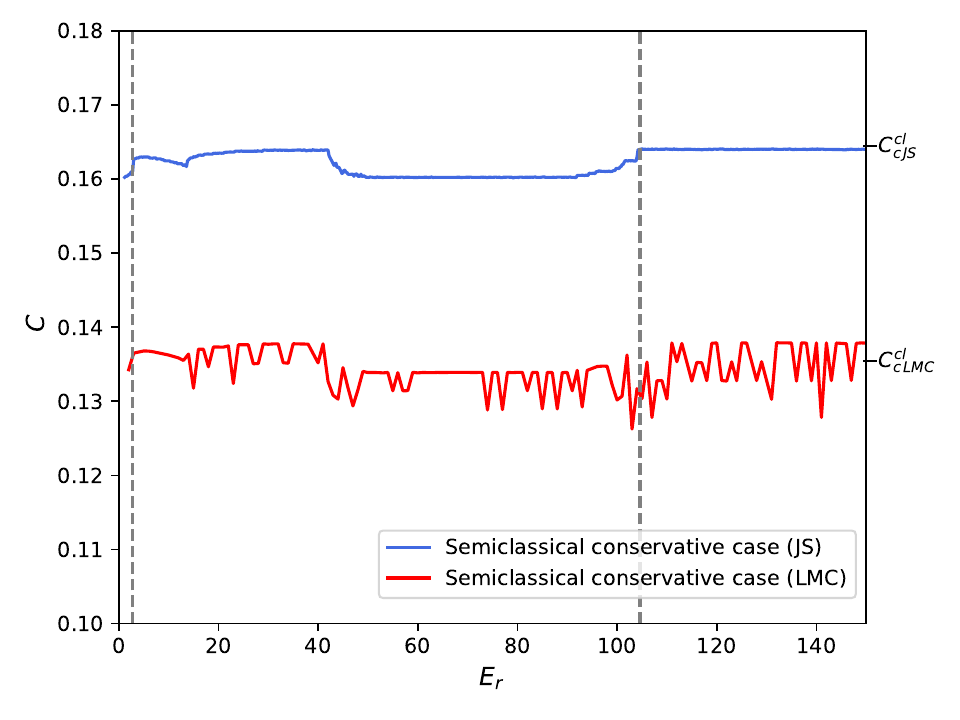}
       \caption{}
     \label{CvsEr_c_JSvsLMC}
    \end{subfigure}
      \centering
      \begin{subfigure}{0.49\textwidth}
        \centering
        \includegraphics[height=6cm]{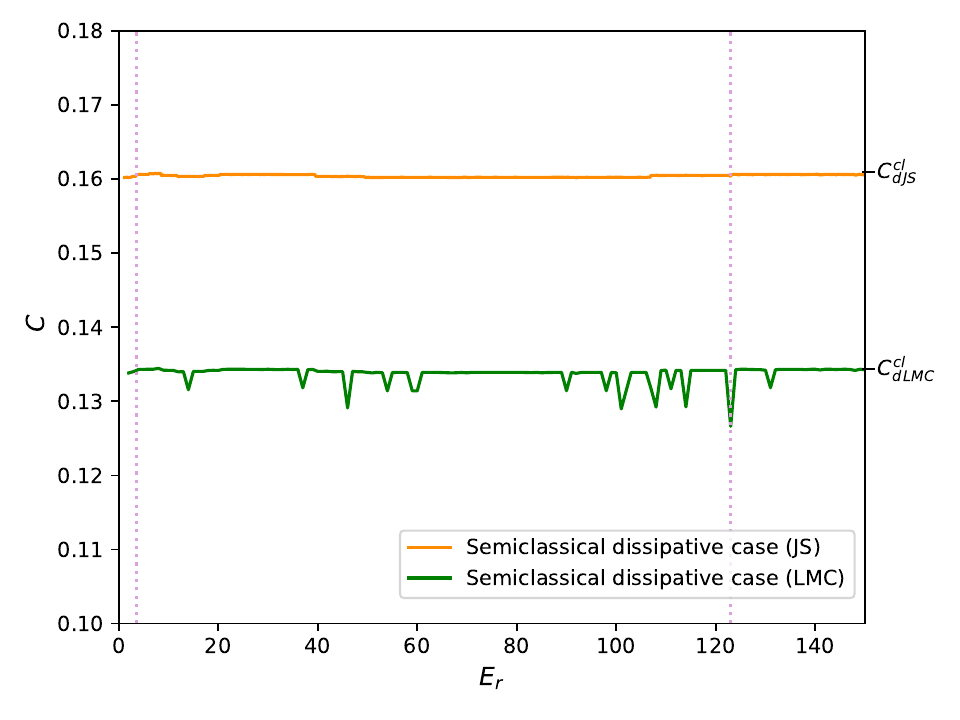}
       \caption{}
        \label{CvsEr_d_JSvsLMC}
    \end{subfigure}
    \caption{Comparative  figure between $C_{LMC}$ and $C_{JS}$ as a function of the relative energy $E_r$. (a) Conservative dynamics. The blue  and red curves exhibit the Jensen-Shannon and Lopez-Ruiz et al. statistical complexities, respectively. The gray dashed vertical lines indicate the transition zone (between $2.8<E_r<104.8$). (b) Dissipative case. The orange and green curves display the Jensen-Shannon and Lopez-Ruiz et al. complexities, respectively. The plum colored vertical dashed lines indicate the transition zone ($3.6 < E_{r_0 }< 123.0$). In addition to the fluctuations of $C_{LMC}$, these plots show a difference in scale between both complexities. This is due to the differences between the normalization of the Jensen-Shannon divergence and the Lopez-Ruiz et al. disequilibrium.}
\end{figure}

\begin{figure}[h]
    \centering
      \begin{subfigure}{0.49\textwidth}
        \centering
        \includegraphics[height=6cm]{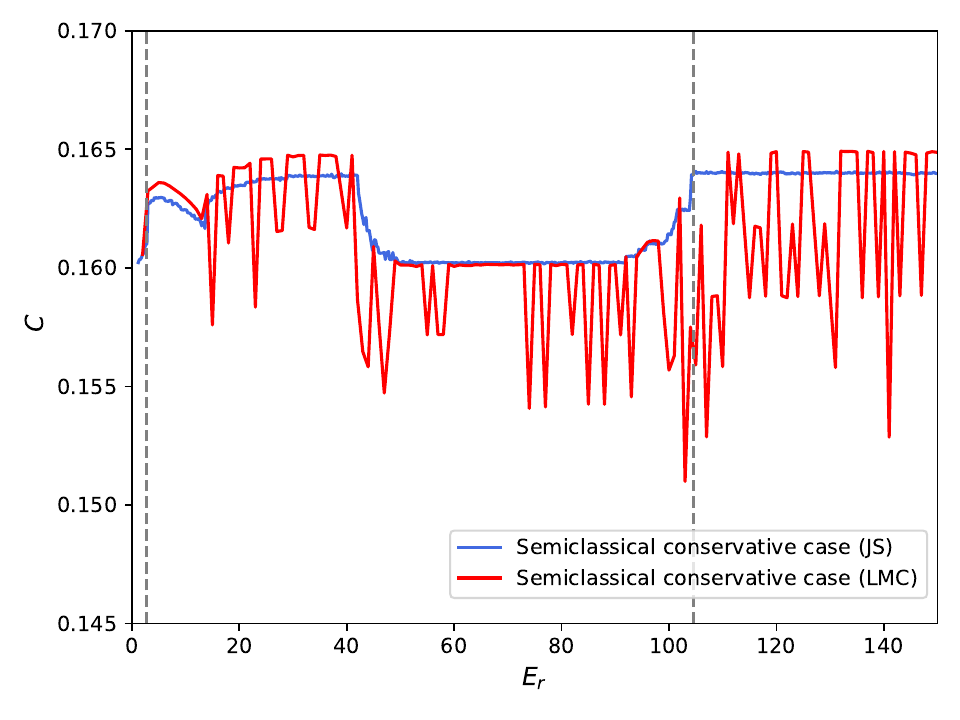}
       \caption{}
       \label{CvsEr_dif_c_JSvsLMC}
    \end{subfigure}
      \centering
      \begin{subfigure}{0.49\textwidth}
        \centering
        \includegraphics[height=6cm]{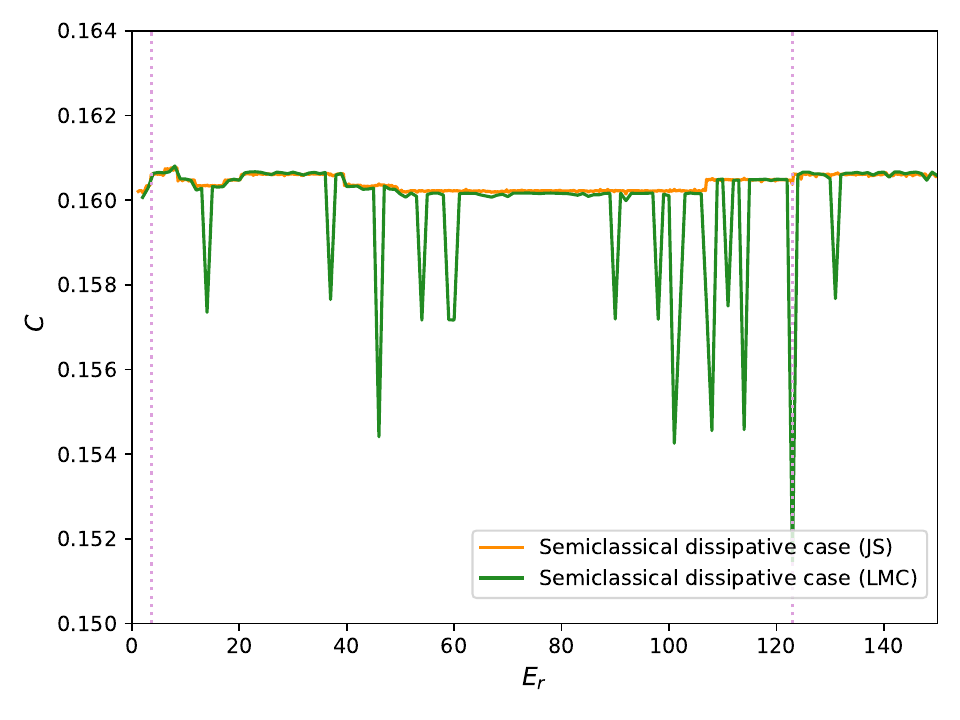}
       \caption{}
        \label{CvsEr_dif_d_JSvsLMC}
    \end{subfigure}
    \caption{Comparative  figure between $C_{LMC}$ and $C_{JS}$ vs. $E_r$. In this case we have changed the scale of $C_{LMC}$ by a factor of $1.196$.  (a) Conservative regime and (b) dissipative one.}
\end{figure}

Figures \ref{CvsEr_c_JSvsLMC} and \ref{CvsEr_d_JSvsLMC} are comparative graphs between $C_{LMC}$ and $C_{JS}$ for both regimes.  The behavior of the curve of $C_{LMC}$ is more erratic that the  $C_{JS}$ one (due to $D_{LMC\, c}$ and $D_{LMC\, d}$ behaviours). This happens for both the conservative and dissipative systems.
On the other hand, the different criteria used to normalize the disequilibrium in the expressions of $C_{LMC}$ and $C_{JS}$ cause a difference in scale.  Figures \ref{CvsEr_dif_c_JSvsLMC}
and \ref{CvsEr_dif_d_JSvsLMC}, expose this characteristic for the conservative and dissipative regimes, respectively. To achieve a meaningful comparison, it was convenient to apply a translation to the complexity $C_{LMC}$ by multiplying all values of the curve by the factor $1.196$. It is observed that, despite the marked fluctuations in the amplitude of  $C_{LMC}$ compared to  $C_{JS}$, both exhibit a similar profile. This phenomenon highlights the consistency in the underlying behavioral pattern.

\begin{figure}[h]
    \centering
      \begin{subfigure}{0.45\textwidth}
        \centering
        \includegraphics[height=4.5cm]{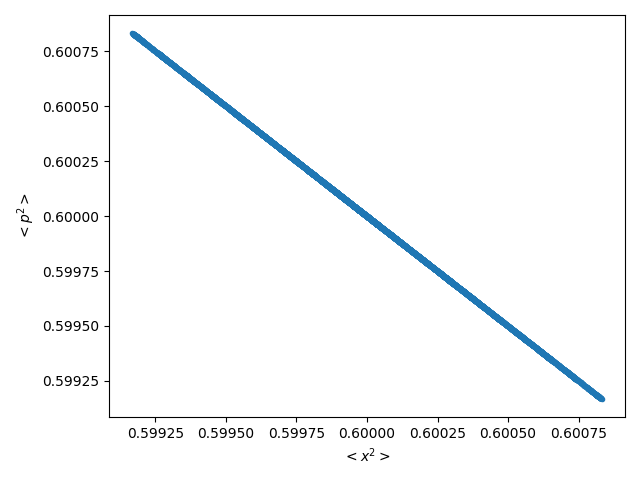}
        \caption{$E_{r}=1.000001$ (quasi-quantum zone).}
    \end{subfigure}
      \centering
      \begin{subfigure}{0.45\textwidth}
        \centering
        \includegraphics[height=4.5cm]{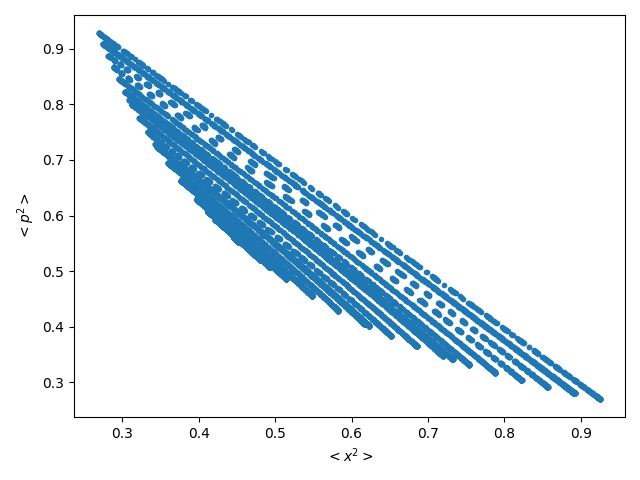}
        \caption{$E_{r}=1.02$ (quasi-quantum zone)).}
    \end{subfigure}
    
    \begin{subfigure}{0.45\textwidth}
        \centering
        \includegraphics[height=4.5cm]{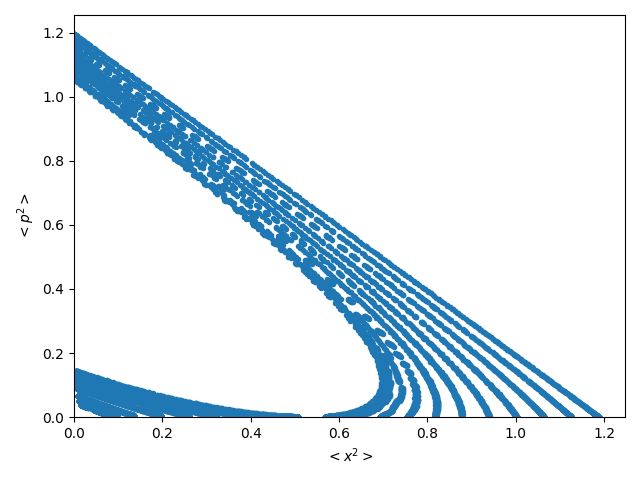}
        \caption{$E_{r}=24.24$ (transitional (mezoscopic) zone).}
    \end{subfigure}
    \begin{subfigure}{0.45\textwidth}
        \centering
        \includegraphics[height=4.5cm]{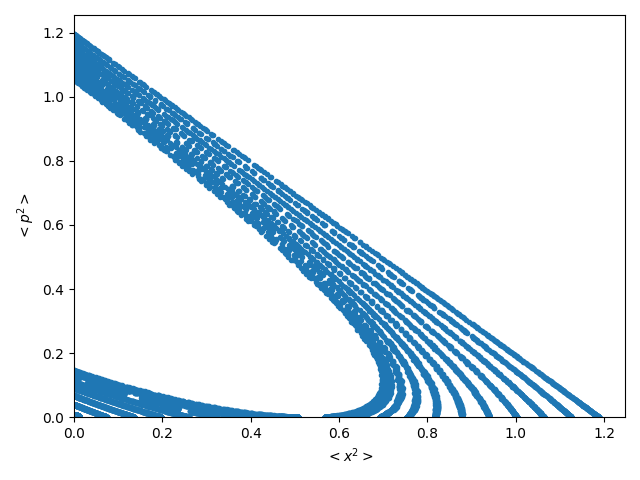}
        \caption{$E_r=104.6$ (critical point when  convergence begins).}
    \end{subfigure}
    
    \begin{subfigure}{0.45\textwidth}
        \centering
        \includegraphics[height=4.5cm]{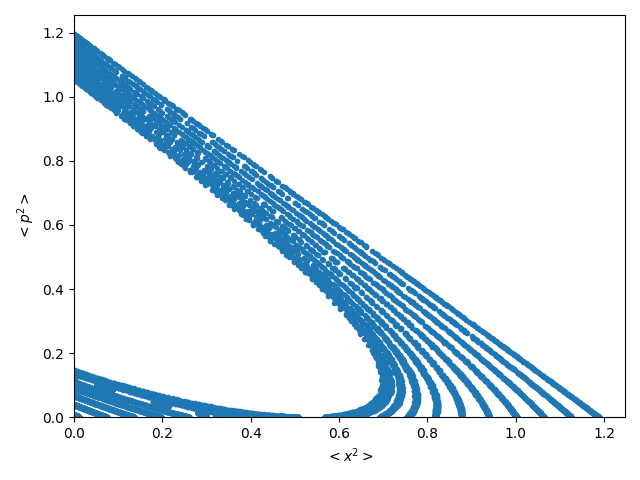}
        \caption{$E_r=4\cdot 10^{4}$  (classical zone).}
    \end{subfigure}
    \begin{subfigure}{0.45\textwidth}
        \centering
        \includegraphics[height=4.5cm]{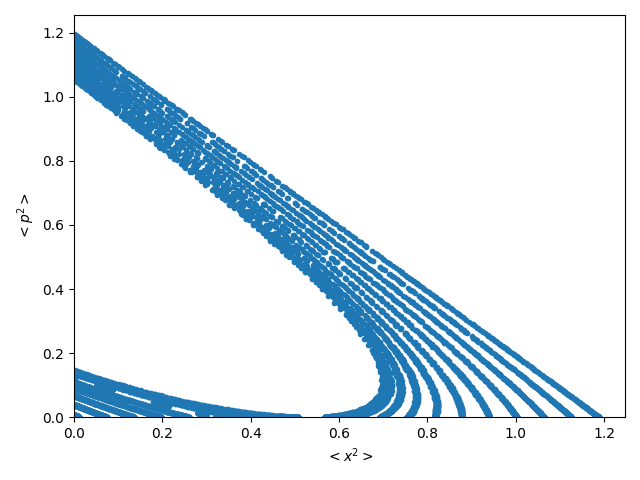}
        \caption{$E_r=\infty$ (classical analogous). }
    \end{subfigure}
    \caption{Poincaré sections corresponding to the conservative dynamics.}
    \label{PScons}
\end{figure}

\begin{figure}[h]
    \centering
     \begin{subfigure}{0.45\textwidth}
        \centering
        \includegraphics[height=4.5cm]{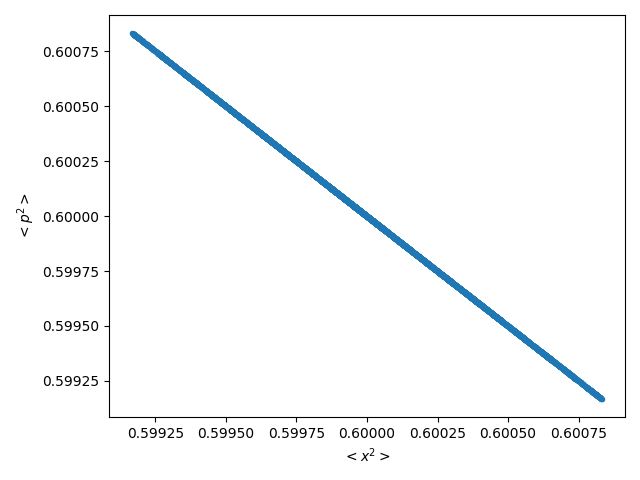}
        \caption{$E_r=1.000001$ (quasi-quantum zone).}
    \end{subfigure}
    \centering
     \begin{subfigure}{0.45\textwidth}
        \centering
        \includegraphics[height=4.5cm]{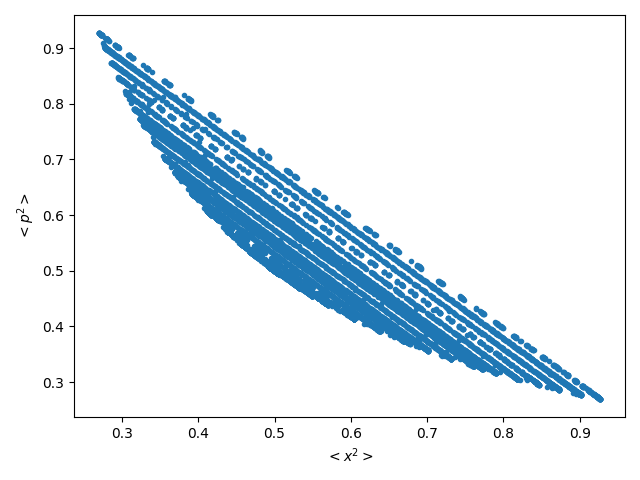}
        \caption{$E_r=1.02$ (quasi-quantum zone).}
    \end{subfigure}
    
    \begin{subfigure}{0.45\textwidth}
        \centering
        \includegraphics[height=4.5cm]{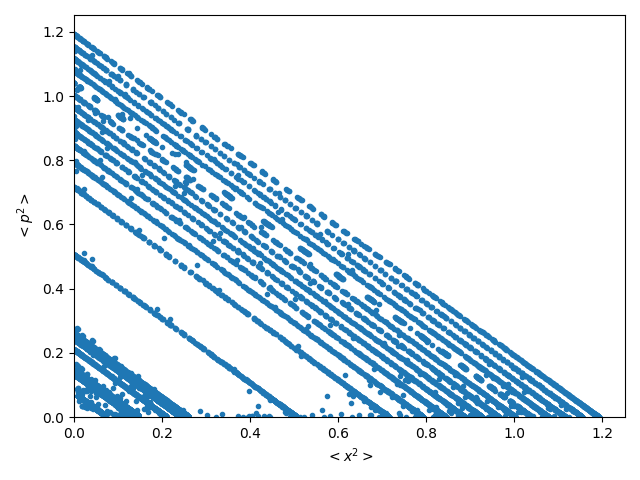}
        \caption{$E_r=24.24$ (transitional (mezoscopic) zone).} 
    \end{subfigure}
    \begin{subfigure}{0.45\textwidth}
        \centering
        \includegraphics[height=4.5cm]{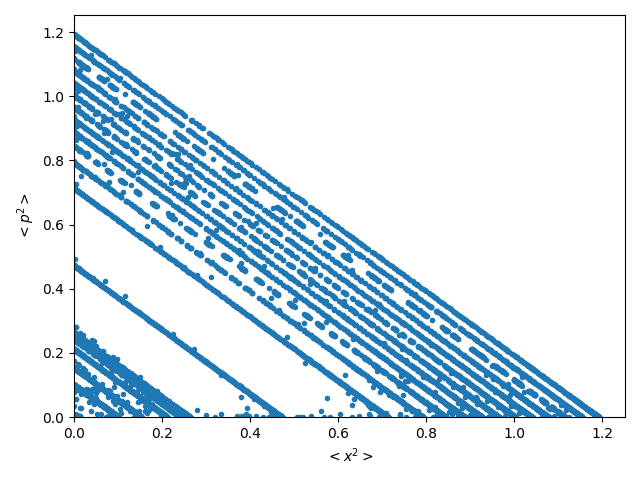}
        \caption{$E_r=123.0$ (critical point when  convergence begins).}
    \end{subfigure}
    
    \begin{subfigure}{0.45\textwidth}
        \centering
        \includegraphics[height=4.5cm]{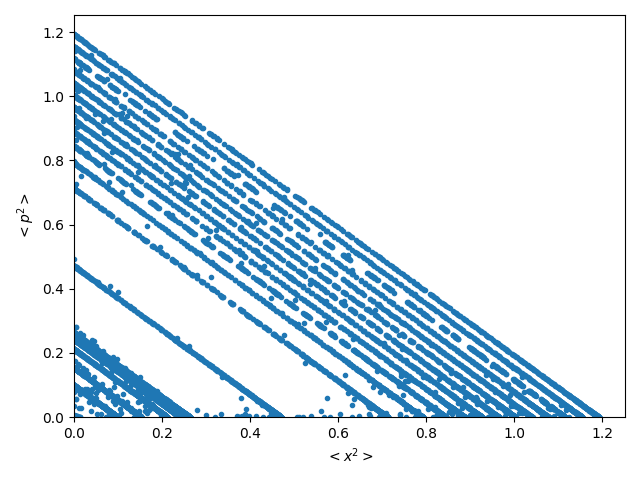}
        \caption{$E_r=4\cdot 10^{4}$  (classical zone).}
    \end{subfigure}
    \begin{subfigure}{0.45\textwidth}
        \centering
        \includegraphics[height=4.5cm]{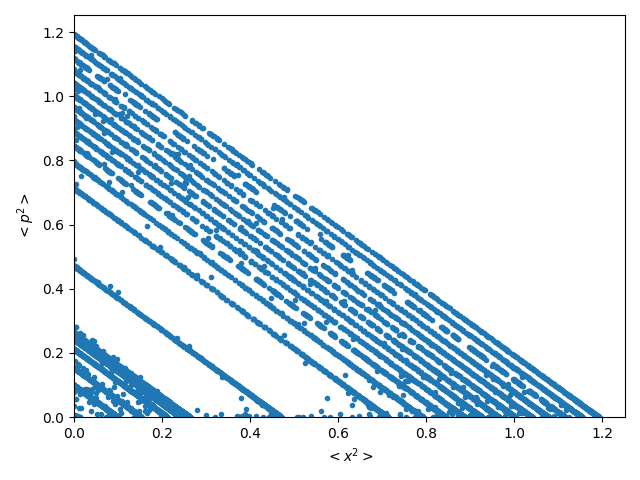}
        \caption{$E_r=\infty$ (classical analogous).}
    \end{subfigure}
    \caption{Projections of the 3d Poincaré sections corresponding to the dissipative case.}
    \label{PSdis}
\end{figure}

In Figures \ref{PScons} and \ref{PSdis} we represent the Poincaré sections (conservative regime) and the projections of the Poincaré sections (dissipative dynamics) (for more details see \cite{Chaos}), suitable for the results obtained in this work.

In Figure \ref{PScons} we plot Poincaré sections corresponding to (a) $E_r=1.000001$ and (b) $E_r=1.02$, both in the semiquantum zone, (c) $E_r= 24.24$ belonging to the transition zone, (d) $E_r=104.6$, the critical point between the transition and classical zones where convergence begins,  (e) $E_r=4\cdot 10^{4}$ in the classical zone. Finally  in (f) we depict the Poincaré section of the conservative classical analogous.   

Figure \ref{PSdis} shows the projections of the 3D dissipative Poincaré sections, for the same $E_r$ values in the same zones as figure \ref{PScons} (\cite{Chaos}), in subfigures (a),  (b),  (c) and (e). Subfigure (d) correspond to the critical point for the disipative regime $E_r=123.0$.  In (f) we plot the projection of the 3D Poincaré section of the dissipative classical analogous.

It is important to observe in these figures the difficulty in precisely defining, using Poincaré Sections, the limit between the three zones. This happens because of the  system dynamics is very regular. However, the information quantifiers used in this work offer better differentiation.

 \newpage
 
\clearpage

\section{Conclusions}

In this article we have analyzed the conservative and dissipative dynamics of the semiclassical Hamiltonian (\ref{eq4}), which contains both quantum and classical variables. For this we have used information quantifiers, such as the Shannon entropy $H$ and the Statistical complexity. In the case of the latter, we analyze two versions, the pioneer one introduced by Lopez-Ruiz et al. $C_{LMC}$ \cite{LR1995, LR2001} and the Jensen Shannon complexity $C_{JS}$ \cite{MPR}. The probability distribution to evaluate all these quantities was extracted from the time series using the Bandt and Pompe permutational method \cite{BP2002}. A methodology that has proven to be robust and reliable.
In particular we have studied the classical limit of the system, that is, the convergence to the classical analogue of $H$ represented by Eq. (\ref{Hclasico}) (only with classical variables), as a function of the relative energy $E_r$ (\ref{eq16}), related to the Uncertainty Principle, with $E_{r} \to \infty$.

 In \cite{Chaos} we had studied the same problem, but using dynamic tools such as Poincaré sections and projections of 3D Poincaré sections. The result found was the existence of three zones, quasi-quantum,  transitional and finally one where the convergence to the results corresponding to the classical analogue was observed (classical zone). However, the boundaries between zones were diffuse, because the dynamics of the system is regular (see figures 5 and 6).
The regularity of the dynamics is reflected in the figures of the present article, by their low values of all information quantifiers. Moreover, after the jump observed in the quasi-quantum zone, one finds a very small band of variation of the respective values. 

It is seen in Figs. 1, 2, 3 and 4 that the Shannon permutation entropy $H$ and the statistical complexities $C_{LMC}$ and $C_{JS}$ confirm the mentioned three zones of the process (quasi-quantum, transitional or mesoscopic and that of decoherence, convergent to the classical analogous result). The corresponding intervals of $E_r$ are the same for the three quantifiers and can be synsetized by the transitional zone. In the conservative scenario $2.8<E_r<104.8$ and in the dissipative regime $3.6<E_r<123.0$.

We have found that the figures of $H$ and $C_{JS}$ have similar morphology but with a difference in scale. This is because the disequilibrium remains almost constant for all values of $E_r$ in both scenarios, conservative and dissipative. Disequilibrium corresponding to the  $C_{LMC}$ complexity is also quasi-constant, but it has greater fluctuations around its mean value (conservative and dissipative regimes).

We conclude that the information quantifiers used in this work provide more details about the convergence process of the semiclassical system to the classical analogous, confirming the results obtained with the Poincaré sections used in Ref. \cite{Chaos}, but with greater precision.

\appendix
\section{Appendix}
\subsection{Semiclassical Model} 

\renewcommand{\theequation}{A\arabic{equation}}
\setcounter{equation}{0} 

Our semiclassical Hamiltonian is (\ref{eq4}). In our methodology, the time evolution of all operator is the canonical one. The Hamiltonian also depends upon the classical variables $A$ and $P_{A}$, which represent the reservoir. These classical quantities obey Hamilton's  equations, of course,  where the temporal generator is  the mean value of the Hamiltonian \cite{Kowalski2002,Cooper1998}. If we add to the system an appropriate ad hoc term we get dissipation \cite{ref,AK1,AK2}.\\
Thus, for a given operator  we have 
\begin{equation}
\dfrac{d \hat{O}}{dt}=\dfrac{i}{\hbar}\left[\hat{H}, \hat{O}\right].
\label{eq1}
\end{equation} 
The evolution of  $\langle \hat{O}\rangle\equiv \textrm{Tr}\left[\rho \hat{O}\left(t\right)\right]$ is provided by  
\begin{equation}
\dfrac{d\langle \hat{O}\rangle}{dt}=\dfrac{i}{\hbar}\left\langle \left[\hat{H},\hat{O}\right]\right\rangle,
\label{eq2}
\end{equation}
using a suitable density operator  $\rho\left(0\right)$. \\
$[\: \hat H, \hat O_{i} \:]$  can always be cast as \cite{Levine,Levine2}
\begin{equation}
\label{conm}
[\: \hat H, \hat O_{i} \:] = i\hbar \sum_{j=1}^{q}g_{ji} \hat
O_{j} \;,
\hskip 1.0cm i = 0, 1, \ldots, q,
\end{equation}
where we can have $q \rightarrow \infty$. We are interested in finite $q$-values. 
In this case it is said that the set of operators $\hat O_{i}, \,\, i = 0, 1, \ldots, q$,  close a Lie semialgebra with $\hat{H}$ \cite{Levine,Levine2}. 
For a semiclassical Hamiltonian like (\ref{eq4}), the coefficients $g_{ji}$ depend on $A$ and $P_A$.
For classical  variables one
\begin{subequations}
\begin{align}
&\dfrac{dA}{dt} =  \dfrac{\partial \langle \hat{H}\rangle }{\partial P_{A}},\\
&\dfrac{dP_{A}}{dt} = - \dfrac{\partial \langle\hat{H} \rangle }{\partial A}-\eta P_{A},
 \end{align}
  \label{eq3}
\end{subequations} 
where if  $ \eta>0$ our system is dissipative. Instead, if  $\eta=0$, it is conservative.\\
To verify these features  \cite{ref,AK1,AK2} a particularly convenient space is used. We will call it ``$v$-space''. With respect to it, a bunch of equations deduced from Eqs. (\ref{eq2}) and (\ref{eq3}) conform an autonomous set of equations. They are first-order coupled differential ones, of the type  \cite{ref,AK1,AK2}
\begin{equation}
\frac{d\vec{v}}{dt} = \vec{F}(\vec{v}),
\label{du}
\end{equation}
with  $ \vec{v} $ is a  ``vector''. It is regarded as a variable containing both classical and quantum parts. We consider volume elements $ V_S $ of this space, surrounded by a surface $S$. The dissipative $\eta$ term generates a contraction of $ V_S $ \cite{Arnold}. The divergence of our vector is seen to be $-\eta$, given the fact that the matrix $G$ in equations (\ref{conm}) is traceless. This happens due  to the canonical character of Eqs. (\ref{eq1}). Consequently. we must have \cite{ref,AK1,AK2}
\begin{equation}
\frac{dV_{S}(t)}{dt} = - \eta V_{S}(t).
\label{dV}
\end {equation}
Accordingly, we deal with  a dissipative system  \cite{Arnold}. If the classical Hamiltonian is of the  form
\begin{equation}
\label{Hcl}
H_{cl} = \frac{1}{2M} P_{A}^{2} + V(A),
\end{equation}
then  the time-evolution of the  total energy $\langle \hat H \rangle$ becomes
\begin{equation}
\frac{d\langle \hat H \rangle}{dt} = - \frac{\eta}{M} P_{A}^{2}.
\label{Energtot}
\end{equation} 
Its meaning is to be ascertained through the lens  of Eq. (\ref{dV}) \cite{ref,AK1,AK2}. 
\\ 
The complete set of equations (\ref{eq2}) + (\ref{eq3}) constitutes an autonomous set of non-linear coupled first-order ordinary differential equations (ODE). They allow for a dynamical description in which no quantum rules of the sub-quantal system are violated, e.g-, the commutation-relations are trivially conserved at all times, since  the quantum  evolution is the canonical one for an effective time-dependent Hamiltonian ($A$  and $P_A$, play the role of time-dependent parameters of the quantum system).
The initial conditions are determined by a  suitable quantum density operator $\hat \rho$. This happens both in the dissipative and in the conservative instances \cite{RK1,AK2}.

\end{document}